\documentclass[10pt,a4paper]{article}

\usepackage{amsmath,amssymb}
\usepackage{amsthm,latexsym}
\usepackage[labelformat=simple]{subfig}

\usepackage{bm,multirow}        
\usepackage{graphicx}

\graphicspath{{./fig/}}

\usepackage{hyperref}

\begin{document}

\title{Model Order Reduction in Neuroscience}

\author{
B\"ulent Karas\"ozen \thanks{Institute of Applied Mathematics \& Department of Mathematics, Middle East Technical University, Ankara-Turkey  {bulent@metu.edu.tr}}}

\date{}

\maketitle

\begin{abstract}
Human brain contains approximately $10^9$ neurons, each  with approximately $10^3$ connections, synapses, with other neurons. Most sensory, cognitive and motor functions of our brains depend on the interaction of a large population of neurons. In recent years, many technologies are developed for recording  large numbers of neurons  either sequentially or simultaneously. Increase in computational power and algorithmic developments have enabled advanced analyses of neuronal population parallel to the  rapid growth of quantity and complexity of the recorded neuronal activity.   Recent studies made use of  dimensionality and model order reduction techniques to extract coherent features which  are not apparent at the level of individual neurons. It has been observed that the neuronal activity evolves on low-dimensional subspaces. The aim of  model reduction of large-scale neuronal networks is accurate and fast prediction of  patterns and their propagation in different areas of brain. Spatiotemporal features of the brain activity  are identified on low dimensional subspaces with methods such as
dynamic mode decomposition (DMD), proper orthogonal decomposition (POD), discrete empirical interpolation (DEIM) and combined parameter and state reduction.
In this paper, we give an overview about the currently used dimensionality reduction and model order reduction techniques  in neuroscience.\\


\noindent\textbf{\textit{Classification[MSC~2010]:}}  93A15,92C55, 37M10,37M99,37N40,65R32.
\end{abstract}

\section{Introduction}

Due to the advances in recording and imaging technologies,  the number of recorded  signals from brain cells increased significantly in the last few years. The recorded spatio-temporal neural activity give rise to  networks with complex dynamics. In neuroscience,
molecular and cellular
level details are incorporated in large-scale models of the brain in order to
reproduce phenomena such as learning and behavior.
The rapid growth of simultaneous neuronal recordings  in scale and resolution brings challenges with the analysis of the neuronal population activity. New computational approaches have to  be  developed to analyze, visualize, and understand large-scale recordings of neural activity. While algorithmic developments and the availability of significantly more computing power  have enabled analysis of larger neuronal networks, these advances cannot keep pace with increasing size and complexity of recorded activity.
The activity of complex networks of neurons can often be described by relatively few distinct patterns. Model order reduction techniques enable us  to identify the coherent spatial–temporal patterns. \\

The presence or absence of a neural mechanism can be analyzed  for neuronal populations. Dimensionality reduction methods \cite{Cunningham14}  are data-driven statistical techniques for forming and evaluating hypotheses about population activity structure, which are summarized in Section 2.
One of the goals of neurosciences is fast and accurate predictions of
the potential propagation in neurons. The differential equations describing the propagation of potential
in neurons were developed  and are described by  Hodgkin and Huxley equations \cite{Hodgkin90}. They consists of a coupled system of ordinary
 and partial  differential equations (ODEs and PDEs). The dimension of the
associated discretized systems is very large for accurately simulating neurons with realistic morphological structure and synaptic inputs.
In Section 3 we present two model order reduction approaches based on POD and DEIM \cite{Chaturantabut10} which  can predict accurately the potential propagation in large scale neuronal networks leading to important speedups \cite{Kellems09,Kellems10,Amsallem16}. Using the  functional neuroimagining data  from
electroencephalography (EEG) or  functional magnetic resonance
imaging (fMRI),
different regions of the brain can be inferred  by dynamic causal modeling (DCM) \cite{Friston03}.
Effective connectivity is parameterised in terms of
coupling among unobserved brain states and neuronal activity
in different regions. In Section 4 we describe a  combined state and parameter reduction for parameter estimation and identification \cite{Himpe15} to extract effective connectivity in neuronal networks from measured data,
such as data from electroencephalography (EEG) or functional magnetic resonance
imaging (fMRI). In Section 5 the data-driven, equation free, model order reduction method dynamic mode decomposition (DMD)  is described for identifying  sleep spindle networks \cite{Brunton16}. Reduced order models with POD and DEIM and four variants of them are presented for neuronal synaptic plasticity and neuronal spiking networks in Section 6.

\section{Dimensionality reduction methods}
Coordination of responses across neurons exists only at the level of the population and not at the level of single neurons.  The presence or absence of a neural mechanism can be analyzed  for neuronal populations. Dimensionality reduction methods are data-driven statistical techniques for forming and evaluating hypotheses about population activity structure.
They produce low-dimensional representations of high-dimensional data with the aim to extract coherent patterns which preserve  or highlight some feature of interest in the data \cite{Cunningham14}.   The recorded neurons of dimension $D$ are likely not independent of each other, because they belong to a common network of neuronal populations. From the high-dimensional data  of neuronal recordings, a smaller number of explanatory variables $K$ ( $K < D$) are extracted with the help of dimensionality reduction methods. The explanatory variables are not directly observed, therefore they are referred to as latent variables. The latent variables define a $K$-dimensional space representing  coherent patterns of the noisy neural activity of $D$ neurons.\\

There exists several dimensionality reduction methods which differ in the statistical interpretation of the preserved and discarded features of the neuronal populations. We summarize the commonly used statistical methods of dimensionality reduction methods in \cite{Cunningham14}, where further references about the methods can be found.\\

{\em Principal component and factor analysis}; The most widely known technique to extract coherent patterns from high dimensional data is the modal decomposition.  A particularly popular modal decomposition technique is principal component analysis (PCA), which derives modes ordered by their ability to account for energy or variance in the data. In particular, PCA is a static technique and does not model temporal dynamics of time-series data explicitly, so it often performs poorly in reproducing dynamic data, such as recordings of neural activity. The low-dimensional space identified by PCA captures variance of all types, including firing rate variability and spiking variability, whereas factor analysis (FA)  discards the independent variance for each neuron.  and preserves variance that is shared across neurons.\\

{\em Time series methods}: The temporal dynamics of the population activity can be identified if the data comes  from a time series. The most commonly used time series methods for dimensionality reduction neural recordings are: hidden Markov models (HMM) \cite{Kemere08},  kernel smoothing followed by a static dimensionality reduction method, Gaussian process factor analysis (GPFA) \cite{Yu09}, latent linear dynamical systems (LDS)  \cite{Buesing12} and latent nonlinear dynamical systems (NLDS) \cite{Patreska11}. They produce latent neural trajectories that capture the shared variability across neurons. The HMM is applied when a jump between discrete states of neurons exists, other methods identify smooth changes in firing rates over time. \\

{\em Methods with dependent variables}: On many neuronal recordings the high-dimensional firing rate space is associated with labels of one or more dependent variables, like stimulus identity, decision identity or a time index. The dimensionality reduction aims in this case to project the data such that differences in these dependent variables are preserved. The linear discriminant analysis (LDA) can be used to find a low-dimensional projection in which the $G$ groups to which the data points belong are well separated. \\

{\em Nonlinear dimensionality reduction methods}:
All the previous methods assume a linear relationship between the latent and observed variables. When the data  lies on a low-dimensional, nonlinear manifold in the high-dimensional space, a linear method may require more latent variables than the number of true dimensions of the data. The most frequently used  methods to identify nonlinear manifolds are Isomap79 \cite{Tenenbaum00}  and locally linear embedding (LLE) \cite{Roweis00}. Because the nonlinear methods use local neighborhoods to estimate the structure of the manifold, population responses may not evenly explore the high-dimensional space. Therefore theses methods should be used with care.\\

\section{Proper orthogonal decomposition (POD) and discrete empirical interpolation (DEIM) for Hodgin-Huxley model}

One of the goals of neurosciences is fast and accurate predictions of
the potential propagation in neurons. The differential equations describing propagation of potential
in neurons are described by  Hodgkin and Huxley (HH) cable equations \cite{Hodgkin90}. They consists of a coupled system of ordinary
(ODEs) and partial  differential equations PDEs.  Accurate simulation of morphology, kinetics and synaptic inputs of  neurons requires solution of large systems of nonlinear ODEs.
The complexity of the models are determined by the synapse density of the dentritic length ($1\mu$ one micron). In simulations, for one synapse per micron on a cell $5$ mm dendrite at $5,000$ compartments each with $10$ variables are needed, which results in $50,000$ coupled nonlinear system of ODEs \cite{Kellems09,Kellems10}.
To recover complex dynamics, efficient reduced order neuronal methods are developed using the POD and DEIM from the snapshots of the in space and time discretized coupled PDEs and ODEs \cite{Kellems09,Kellems10,Amsallem16}. In this section we describe two of them. They differ in the formulation of the HH cable equation and of the equations for the gating variables.
\subsection{Morphologically accurate reduced order modeling}

The neuronal full order models (FOMs) in \cite{Kellems09,Kellems10} consists of $D$ branched dendritic neurons  $B = \sum_{d=1}^D B_d$  meeting at the soma, where the $d^{th}$
has $B_d$ branches. It is assumed  that the branch $b$ carries $C$ distinct ionic currents with
associated densities  and $G_{bc}(x)$ and reversal potentials
$E_c, c = 1,\ldots,C$. The kinetics of current $c$ on
branch $b$ is  governed by the $F_c$ gating
variables, $w_{bcf} , f = 1, \ldots , F_c$. When subjected to input
at $S_b$ synapses, the nonlinear HH cable equation for the  transmembrane potential $v_b(x,t)$  with the equation for the gating variables
$ w_{bcf}$ is given by (see \cite{Amsallem16} Fig.1, model network with three cables)
\begin{eqnarray} \label{hh3}
a_bC_m\frac{\partial v_b}{\partial t}  =  & & \frac{1}{2R_i}\frac{\partial }{x}\left (a_b^2 \frac{\partial v_b}{\partial x} \right ) \nonumber \\
&   & -a_b  \sum_{c=1}^C G_{bc}(x)  (v_b -E_c) \prod_{f=1}^{F_c} w_{bcf}^{q_{cf}} \nonumber \\
 &  & \frac{1}{2\pi} \sum_{s=1}^{S_b} g_{bs}(t)\delta (x- x_{bs}) (v_b -E_{bs}) \\
\frac{\partial w_{bcf}}{\partial t} & = & \frac{w_{cf,\infty}(v_b) - w_{bcf}}{\tau_{cf}(v_b)},\quad 0 < x < l_b, \; t >0,
\end{eqnarray}
where $g_{bs} (nS)$ is the time course, $x_{bs}$ is the spatial location and $E_{bs}$ is the reversal potential of the $s$th  synapse  on branch $b$.
The variables and parameters in \eqref{hh3} are described in \cite{Kellems09,Kellems10}.

These branch potentials interact at $J$ junction points,
where junction $J$ denotes the soma. The $D$ dendrites join at soma. Continuity of the potential at the soma leads to a common value at current balance  denoted by $v_\sigma (t)$. Then the networked form of \eqref{hh3}
becomes
\begin{eqnarray} \label{hh4}
a_bC_m\frac{\partial v_\sigma}{\partial t}  =  &  & \frac{\pi}{A_\sigma R_i}
\sum_{d=1}^D \frac{\partial }{\partial x}\left ( a_{b_J^d}^2 (l_{b_J^d})\frac{\partial v_{b_{J^d}} (l_{b_{J^d}},t)}{\partial x} \right ) \nonumber \\
 &  & -a_b\sum_{c=1}^C G_{\sigma c}(x)  (v_\sigma -E_c) \prod_{f=1}^{F_c} w_{\sigma cf}^{q_{cf}} (t) \nonumber \\
   &  & \frac{1}{A_\sigma}\sum_{s=1}^{S_b} g_{\sigma s}(t) (v_\sigma(t) -E_{\sigma s}) \\
\frac{\partial w_{\sigma cf}(t)}{\partial t} & =  & \frac{w_{cf,\infty}(v_\sigma(t)) - w_{\sigma cf}(t)}{\tau_{cf}(v_\sigma)(t)}, \quad 0 < x < l_b, \; t >0.
\end{eqnarray}

When the cell is partitioned into $N$ compartments,
with $C$ distinct ionic currents per compartment and with $F$ gating variables per current,
the following nonlinear ODEs are obtained
\begin{eqnarray} \label{hhdis}
v'(t)  =   & &  Hv(t) -(\Phi(w(t)e). v(t) + \Phi (w(t))E_i \nonumber \\
 & &  + G(t).(v(t)-E_s), \quad   v(t) \in \mathbb{R}^N\\
w'(t)   = & &  (A(v(t)) -w(t))./B(v(t)), \quad w(t) \in \mathbb{R}^{N\times C \times F}
\end{eqnarray}
where
$H\in \mathbb{R}^{N \times N}$ is the Hines matrix \cite{Hines84}, $e=[1\;1\cdots 1]^T\in \mathbb{R}^C$ and the ‘dot’ operator, $a.b$, denotes element-wise multiplication. $E_i$ and $E_s$  are respectively the vector of channel reversal potentials and  the vector of synaptic reversal potentials, respectively
Eq. \eqref{hhdis} is discretized in time by the second order discretized Euler scheme \cite{Hines84}. \\

In \cite{Kellems10} using the snapshots of $v(t)$ and of the nonlinear term $N(v(t),w(t)) \equiv (\Phi)w(t))e).v(t) - \Phi(w(t)))E_i$ at times $t_1,t_2,\ldots,t_n$ the POD and DEIM modes are constructed.

The reduced membrane potential $v_r$ are constructed using the POD basis, the reduced gating variables $w_r$ are obtained after applying the DEIM to the nonlinear terms.  The reduced order model in \cite{Kellems10} preserves the spatial
precision of synaptic input, captures accurately the subthreshold
and spiking behaviors. \\

In \cite{Kellems09} a linearized quasi active reduced neuronal model is constructed using balanced truncation  and ${\mathcal H}_2$ approximation of transfer functions in time. ROMs preserve the input-output relationship and reproduce only subthreshold dynamics.

\subsection{Energy stable neuronal reduced order  modeling}

In \cite{Amsallem13,Amsallem16} a different form of the HH cable equation and ODEs for gating variables is considered.
The intracellular potential $v(x,t)$ and three
gating variables $m(x,t),\; h(x,t)$, and $n(x,t)$  describe the activation and decativation of the ion channels, of the sodium channels and of  the  potassium channels, respectively. A single cable in the
computational domain $(x,t) \in [0,L] \times (0,T]$ describing the distribution of the potential $u(x,t)$ is given by \cite{Amsallem13,Amsallem16}

\begin{equation} \label{hh1}
\frac{\partial u}{\partial t} = \frac{\mu }{a(x)} \left ( a(x)^2u_x\right )_x - \frac{1}{C_m}g(m,h,n)u + \frac{1}{C_m}f(m,h,n,x,t),
\end{equation}
where $a(x)$ the radius of the neurons and $C_m$ is specific membrane capacitance, $\mu = \frac{1}{2C_mR_i} > 0$ the ratio with $R_i$ the axial resistivity.
The conductance $g(x,t)$ is a polynomial of the gating variables
\begin{equation}
g(x,t) = g_1m^3h+g_2n^4 +g_3 > 0,
\end{equation}
with the source term
\begin{equation}
f(m,h,n,x,t) = g_1E_1m^2h + g_2E_2n^4 + g_3E_3 - i(x,t),
\end{equation}
where $E_l, \;l=1,2,3$ are equilibrium potentials and $i(x,t)$ input current at $x$
\begin{equation}
i(x,t) = \sum_{s=1}^{N_s} i_s(x,t), \quad x \in [0,L].
\end{equation}
The nonlinear ODEs for the gating variables are given by
\begin{eqnarray}\label{gating1}
\frac{\partial m }{\partial t} &  = &  \alpha_m(v(x,t)(1-m(x,t)) -\beta_m v(x,t)) m(x,t),   \nonumber \\
\frac{\partial h }{\partial t} & =  & \alpha_h(v(x,t))(1-h(x,t)) -\beta_h v(x,t)) h(x,t), \\
\frac{\partial n }{\partial t} &  =  & \alpha_n(v(x,t))(1-n(x,t)) -\beta_n v(x,t)) n(x,t),  \nonumber
\end{eqnarray}
Expression for $\alpha_m, \; \alpha_h, \; \alpha_n, \;  \beta_m, \; \beta_h, \; \beta_n$ and boundary conditions can be found in \cite{Amsallem16}.

In \cite{Amsallem13,Amsallem16}, a model network with three cables connected to a soma is used. The equations governing the potential propagation in the network $N_c$ neuron cables-dentrites and /or axons with the superscript $^{(c)},\; c=1,\ldots N_c$, are given as
\begin{eqnarray}\label{hh2}
\frac{\partial v^{(c)}}{\partial t}  =  &  &\frac{\mu }{a^{(c)}(x^{(c)})}
\left( \left( a^{(c)}\left( x^{(c)}\right)^2
\right)v^{(c)}_x
\right)_x - \frac{1}{C_m}g\left(m^{(c)},h^{(c)},n^{(c)}\right)u^{(c)} \nonumber\\
+ & & \frac{1}{C_m}f\left(m^{(c)},h^{(c)},n^{(c)},x^{(c)},t\right)
\end{eqnarray}

\begin{eqnarray}\label{gating2}
\frac{\partial m^{(c)} }{\partial t} &  = &  \alpha_m(v^{(c)}(1-m^{(c)}) -\beta_m v^{(c)}) m^{(c)},   \nonumber \\
\frac{\partial h^{(c)} }{\partial t} & =  & \alpha_h(v^{(c)})(1-h^{(c)}) -\beta_h v{(c)}) h^{(c)}, \\
\frac{\partial n^{(c)} }{\partial t} &  =  & \alpha_n(v^{(c)}))(1-n^{(c)}) -\beta_n v^{(c)}) n^{(c)},  \nonumber
\end{eqnarray}
for $x^{(c)} \in \Omega^{(c)} =[0,L^{(c)}]$ together with the boundary conditions.\\

The semi-discrete form of these  equations are approximated using energy stable summation by parts \cite{Amsallem13,Amsallem16} for the model network.
The reduced order bases (ROB) for multiple cables of identical lengths are assembled into
a network in block form.
The block structure of the ROB allows a flexible structure-preserving model
reduction approach with an independent approximation in each cable and  energy
stability and accuracy properties follow from this block structure.
Computation of the time varying reduced variables in the gating equations at every time $t$ is costly because they scale with dimension of FOM.    A nonnegative
variant of the discrete empirical interpolation method (NNDEIM) is developed in \cite{Amsallem16} to preserve the
structure and energy stability properties of the equations.\\


The capability
of the greedy-based approach to generate accurate predictions in large-scale neuronal
networks is demonstrated for system with more than $15,000$ degree of freedoms (dofs).
The state variable ROB of dimension $l = 15$ with POD modes together with the nonnegative ROBs of
dimension $p = 60$ with NNDEIM modes are constructed using a greedy approach to predict
the potential variation at the soma. The speedup of simulations is about $20$ larger than Galerkin
projection only is $1.3$ without using the NDEIM.

\section{Combined state and parameter reduction for dynamic causal modelling}

In neuroscience different regions of the brain are inferred using neuroimagining data from EEG or fMRI recordings
using the method od dynamic causal modeling (DCM) \cite{Friston03}.
Effective connectivity is parameterised in terms of
coupling among unobserved brain states and neuronal activity
in different regions. In DCM the neuronal activity is of the observed brain region is represented as a SISO (single input single output) linear state-space system
\begin{equation}
\dot{x} = A_{\mathrm dyn}(\mu) x + B_{\mathrm dyn}u
\end{equation}
with the parameterized connectivity $A_{\mathrm dyn}(\mu)$ and external input matrices $B_{\mathrm dyn}$.

Linearization of the nonlinear DCM hemodynamic forward sub-model (balloon model) \cite{Friston03} transforms the neuronal activity to the measured BOLD (blood oxygen level dependent) response. Linearization around the equilibrium results in the following single input, single output (SISO) system:

\begin{eqnarray} \label{siso}
B_{obs} & := &  (1 \; 0 \; 0 \; 0)^T,\quad
C_{obs} =   ( 0\; 0 \; V_0k_1 \; V_0k_2),\\
\dot{z}_i & = & A_{obs}z_i + B_{obs}x_i,\\
y_i & = & C_{obs}z_i,\\
z_0 & = & (0 \; 0 \; 0 \; 0)^T,
\end{eqnarray}

\begin{equation}\label{obs}
A_{\mathrm obs}  := \left (  \begin{array}{cccc}
\frac{1}{\tau_s}  & \frac{1}{\tau_f} & 0 & 0 \\
1 & 0 & 0 & 0 \\
0 & \frac{1}{\tau_0 E_0} (1-(1-E_0)(1-\ln(1-E_0))) & \frac{1}{\tau_0} & \frac{1-\alpha}{\tau_0\alpha}  \\
0 & \frac{1}{\tau_0} & 0 &  \frac{1}{\tau_0\alpha}
\end{array} \right ) .
\end{equation}

The fMRI measurements at the $i^{th}$ brain region are reflected by the output variables $y_i$.
For the meaning of the variables and parameters in \eqref{siso}  and \eqref{obs} we refer to \cite{Himpe15,Himpe16}.
The linearized forward sub-models are embedded into the fMRI connectivity model
\begin{equation} \label{joint}
\left ( \begin{array}{c} \dot{x} \\ \dot{z}_1 \\ \dot{z}_2 \\ \vdots \\ z_{N_{dyn}}     \end{array} \right )
=  \left ( \begin{array}{ccccc} A_{dyn}(\mu) & 0 & 0 & \cdots & 0 \\
\delta_{1,1} & A_{obs} & 0 & & \\
 \delta_{2,1} & 0 & A_{obs} &  & \\
 \vdots & & \ddots & \\
 \delta_{1,N_{dyn}} & & & A_{obs}
\end{array} \right )
\left ( \begin{array}{c} x \\ z_1 \\ z_2 \\ \vdots \\ z_{N_{dyn}}     \end{array} \right )
+ \left ( \begin{array}{c}  B_{dyn} \\ 0 \\ 0 \\ \vdots \\ 0     \end{array} \right )v,
\end{equation}

\begin{equation}\label{joint1}
y = \left ( 0 \left ( \begin{array}{ccc} C_{obs} & & \\  & \ddots & \\  & & C_{obs}   \end{array} \right ) \right )
\left ( \begin{array}{c} x \\ z_1 \\ z_2 \\ \vdots \\ z_{N_{dyn}}     \end{array} \right ),
\end{equation}
where $\delta_{ij}\in \mathbb{R}^{4 \times N_{\mathrm dyn}}$ denotes the Kronecker matrix with non-zero elements located at the $(i,j)^{th}$ component. \\

The linearized state-space forward model \eqref{joint} and \eqref{joint1} corresponds to a multiple input, multiple output (MIMO) system

\begin{equation}
\dot{x}(t) = A(\mu)x(t) + Bu(t), \qquad y(t) = Cx(t),
\end{equation}
where $x\in \mathbb{R}^N$ is the internal state, $u\in \mathbb{R}^J$ the  external input,  $y\in \mathbb{R}^O$  the observed output and
$\mu$ are the parameters describing different conditions.

For large number of $M := N^2$ parameters, the computational cost for inferring the parameters and states is very high.  In \cite{Himpe15,Himpe15a} a combined state and parameter  model order reduction is developed for parameter estimation and identification to extract effective connectivity.
The inversion procedure consists of two phases, the offline and online phases.
In the offline phase, the underlying parameterized model is reduced
jointly in states and parameters. In online phase, the reduced order
model’s parameters are estimated to fit the observed experimental data.
Using state and parameter reduction, the computational cost is reduced in the offline phase.  The simultaneous reduction of state and parameter space is based on Galerkin projections with the orthogonal matrices
for the state $V\in\mathbb{R}^{N\times n}$ and for the parameters $P\in \mathbb{R}^{M\times m}$. The reduced model is of lower order $n<<N,\; m<<M$ than the original full order
model. The reduced states $x_r (t)  \in  \mathbb{R}^n$ and the reduced parameters $\mu \in  \mathbb{R}^m$ are computed as
\begin{equation}
\dot{x}_r(t) = A_r(\mu_r)x_r(t) + B_ru(t), \qquad y_r(t) = C_rx(t)
\end{equation}
with a reduced initial condition $x_{r,0} = V^T x_0$ and the reduced components
\begin{eqnarray*}
   \mu_r &=& P^T\mu \in  \mathbb{R}^m, \\
   A_r(\mu_r)&=& V^T A(P\mu_r)V  \in \mathbb{R}^{n\times n}, \\
   B_r&=&V^TB \in \mathbb{R}^{n\times J},\\
   C_r & = &CV  \in \mathbb{R}^{O\times m}.
\end{eqnarray*}

 In the online phase, the optimization based inverse problem  is combined with the reduction of state and parameter space. The inversion is based on generalized data-driven optimization approach to construct the ROMs  in \cite{Lieberman11} and enhanced with the Monte-Carlo method to speed up the computations. The state projection $V \in \mathbb{R}^{ N\times n}$  and parameter projection $P \in \mathbb{R}^{ m\times m}$  are determined iteratively based on a greedy algorithm by maximizing the error between
the high-fidelity original and the low-dimensional reduced model in the Bayesian setting.

Numerical experiments with the DCM model in \cite{Lieberman11} show highly dimensional  neuronal network system is efficiently inverted due to the  short offline durations.
In the offline phase, Monte-Carlo enhanced methods require more than an order of magnitude less offline time
compared to the original and data-misfit enhanced methods. In the online phase the reduced order model has a speedup factor about an order of magnitude more compared to
the full-order inversion. The output error of the data-misfit enhanced method is close to full order method. The output-errors of Monte-Carlo decrease with increasing number of simulation but does not reach the output error of the full order system.
The source code is available in MATLAB \cite{Himpe15a}.

\section{Dynamic mode decomposition}
Dynamic mode decomposition (DMD) is a data-driven, equation free ROM technique \cite{Kutz16}. It was initially developed to reduce the high dimensional dynamic data obtained from experiments and simulations in the fluid mechanics into a small  number of  coupled spatial–temporal modes  \cite{Rowley09,Schmid10}. DMD was applied to explore spatial–temporal patterns in large-scale neuronal recordings in \cite{Brunton16}. DMD can be interpreted as combination of discrete Fourier transform (DFT) in time and principal component analysis (PCA) \cite{Jolliffe05} in space.
Both PCA  and independent component analyses (ICA) \cite{Hyvarienen00}  are  static techniques, which perform poorly in reproducing dynamic data, such as recordings of neural activity.\\

 The data is taken from  electrocorticography (ECoG) recordings. Voltages from $n$ channels of an electrode array sampled every $\Delta t$. These measurements are arranged at snapshot $k$ to the column vectors ${\mathbf x}_k$. The $m$ snapshots in time construct to data matrices shifted in time with $\Delta t$
\begin{equation}
{\mathbf X} = \left [ \begin{array}{cccc}  | & | &  &  |  \\
{\mathbf x}_1 &  {\mathbf x}_2 & \cdots &  {\mathbf x}_{m-1} \\
 | & | &  &  |  \end{array}
\right ], \quad
{\mathbf X}' = \left [ \begin{array}{cccc}  | & | &  &  |  \\
{\mathbf x}_2 &  {\mathbf x}_3 & \cdots &  {\mathbf x}_m \\
 | & | &  &  |
\end{array}
\right ]
\end{equation}
These matrices are assumed to be related linearly in time
\begin{equation}
{\mathbf X}' = {\mathbf A} {\mathbf X}.
\end{equation}
The DMD of the data matrix pair ${\mathbf X}$  and  ${\mathbf X}'$ is given by the eigendecomposition of ${\mathbf A}$ using the singular value decomposition (SVD) of the data matrix ${\mathbf X} = U\Sigma V^*$ by computing the pseudoinverse $
{\mathbf A} \approx  {\mathbf X}' {\mathbf X}^\dag \equiv {\mathbf X}'{\mathbf V}{\mathbf \Sigma}^{-1}{\mathbf U}^*.
$
The spatio-temporal modes are computed by the exact DMD algorithm \cite{Tu14}.

Because DMD does not contain explicit spatial relationship between neighboring measurements, traveling waves occurring in neuronal networks can not be captured well with a few coherent modes.  DMD was also  used as a windowed technique with a temporal window size constrained  by lower bound as for the discrete Fourier transformation (DFT). In contrast to the fluid dynamics where $n >> m$,  in neuroscience the electrode arrays that have tens of channels  $n$ in the recordings with $m$ number of snapshots in the windows data per second, so that $n < m$. The number of singular values $v$ of ${\mathbf X}$  are less than  $n$ and $m-1$, which restricts the maximum number of DMD modes and eigenvalues to $n$. Because of this the dynamics can be captured over $m$ snapshots.  The rank mismatch is resolved  by appending to the snapshot measurements with $h-1$ time-shifted versions of the data matrices. The augmented data matrix ${\mathbf X}_{\mathrm aug}$ is given  as

\begin{equation}
{\mathbf X}_{\mathrm aug} = \left [ \begin{array}{cccc}  | & | &  &  |  \\
{\mathbf x}_1 &  {\mathbf x}_2 & \cdots &  {\mathbf x}_{m-h} \\
 | & | &  &  |  \\
  | & | &  &  |  \\
  {\mathbf x}_2 &  {\mathbf x}_3 & \cdots &  {\mathbf x}_{m-h-1} \\
   | & | &  &  |  \\
    & & \cdots & \\
  | & | &  &  |  \\
  {\mathbf x}_h &  {\mathbf x}_{h+1} & \cdots &  {\mathbf x}_{m-1}\\
     | & | &  &  |  \\
   \end{array}
\right ].
\end{equation}
The augmented matrices ${\mathbf X}_{{\mathrm aug}}$ and ${\mathbf X}'_{{\mathrm aug}}$  are Hankel matrices, which are constant along the skew diagonal,  as in the Eigenvalue Realization Algorithm (ERA) \cite{Juang85}. The number of the stacks $h$ is chosen such that  $hn >2m$. A measure to determined the optimal number of stacks $h$ is the  approximation error
$$
E = \frac{||{\mathbf X} -\hat{\mathbf X}||_F}{||{\mathbf X}||_F}
$$
where $||\cdot||_F$ is the Frobenius norm. The approximation error $E$ is decreasing with increasing number of stacks $h$ and reaches a plateau, so that the DMD accuracy does not significantly increases. \\
DMD is applied  in \cite{Brunton16} as an automated approach to detect and analyze reliably spatial localization and  frequencies of  sleep spindle networks from human ECoG recordings. A MATLAB implementation is available at \url{github.com/bwbrunton/dmd-neuro/}.

\section{Reduced order modeling of biophysical neuronal networks} \label{plasa}
Recently reduced order models  for ODEs
\begin{equation} \label{plas}
\dot{x}(t) = A(t)x(t) + f(x(t)) + B u(t)
\end{equation}
are constructed using POD and DEIM  to investigate input-output behavior of the neuronal networks in the brain \cite{Lehtimaki17,Lehtimaki19}, where  $x(t)$ are state, and $u(t)$ are input variables, respectively. \\

The model in  \cite{Lehtimaki17} is based on the chemical reactions of molecules in synapses, that are the intercellular information transfer points of neurons. The signaling pathways in
striatal synaptic plasticity are modeled in \cite{Kim13}. This model describes how certain molecules, which are  a prerequisite for learning
in the brain, act in synapses. The stoichiometric equations obey the law
of mass action, which leads to a deterministic system of
$44$ ODEs of the form \eqref{plas} . The state $x(t)$ of the control system  \eqref{plas} is a collection of ions,
molecules, and proteins that act in neuronal synapses.  The linear part of \eqref{plas} is sparse, the nonlinearities are quadratic.
The time dependent stimulus $u(t)$ consists of  molecules that are important for neuronal excitability and plasticity, calcium and glutamate. \\

In  \cite{Lehtimaki19},   a  nonlinear biophysical network model is considered, describing
synchronized population bursting behavior of heterogeneous
pyramidal neurons in the brain \cite{Pinsky94}. Neurons communicate by changing their
membrane voltage to create action potentials (spikes),
propagating from cell to cell. Spiking is the fundamental
method of sensory information processing in the brain, and
synchronized spiking is an emergent property of biological
neuronal networks. The ODE system  \eqref{plas} in  \cite{Lehtimaki19}  consists of the states $x(t)$ as  a collection of $50$ neurons, each modeled with $10$ ODEs, leading totally to a system of ODEs of dimension $500$.  Each cell is modeled with Hodgkin-Huxley equations, where each cell has only two compartments (dendrites and soma) and these compartments have different ion channels.  The state  variables $x(t)$
include the voltages of somatic and dendritic compartments, dendritic calcium concentration, synaptic and ion channel gating variables and  the input $u(t)$ is an injected current. Additionally, the soma compartment voltages are coupled to
dentritic compartments of randomly chosen cells.  This networking of the output of cells as input to other cells is key for producing synchronized population behavior. The nonlinearities are Hodgkin-Huxley type, i.e. exponential functions as well as cubic and quartic polynomials. \\

In  \cite{Lehtimaki17}, POD+DEIM was
applied to a data-driven biological model of plasticity in
the brain \eqref{plas}. The ROMs with POD-DEIM reduce significantly the simulation time and error between the original
model and reduced order solutions can be tuned by adjusting the number
POD and DEIM bases independently. When the ROMs
are  trained in a matching time interval of $10000$ seconds, accurate results are obtained. However,
generalizing the reduced model to longer time intervals is challenging, which is characteristic  for all nonlinear models.

In \cite{Lehtimaki19},  the network model \eqref{plas} is reduced with
localized DEIM (LDEIM)  \cite{Pehersdorfer13}, discrete adaptive
POD (DAPOD)  \cite{Tang18,Yang18},  and adaptive DEIM  \cite{Pehersdorfer15}. DEIM and
the variations are used here in combination with POD. ROMs require large number POD and DEIM bases, to accurately simulate the input-output behavior
in the ROMs. In this model, every cell is heterogeneous
in parameters and there are also jump/reset conditions, which are factors
that pose additional challenges to the reduced order methods.  However, the ROMs in \cite{Lehtimaki19}
were able to replicate the emergent
synchronized population activity in the original network
model. DAPOD and ADEIM perform best in
preserving the spiking activity of the original network model.
ADEIM is too slow and does not allow low enough dimensions to
offset the computational costs of online adaptivity. DAPOD
is able to find a lower dimensional POD basis online than
the other methods find offline, but the runtime is close to the
original model.


\bibliographystyle{plain}
\bibliography{morneuro}

\end{document}